\documentclass{article}
\usepackage[preprint]{dep/spconfa4}
\usepackage{graphicx}
\usepackage{amsmath,amssymb}
\usepackage{mathtools}

\usepackage{import}
\usepackage{cite}

\usepackage{balance}

\usepackage[
	all=normal,
	charwidths=tight,charwidthfraction=0.97, 
	leading=tight,leadingfraction=0.95, 
	tracking=tight,trackingfraction=0.995, 
	wordspacing=tight,wordspacingfraction=0.95 
	]{savetrees}

\usepackage{mathrsfs}
\usepackage{relsize}
\usepackage[colorlinks = true,
	linkcolor = black,
	urlcolor  = black,
	citecolor = black,
	anchorcolor = black]{hyperref}

\newcommand{\fig}[1]{\hyperref[fig:#1]{Fig.~\ref*{fig:#1}}}
\newcommand{\eq}[1]{\hyperref[eq:#1]{Eq.~(\ref*{eq:#1})}}
\newcommand{\sect}[1]{\hyperref[sec:#1]{Section~\ref*{sec:#1}}}
\newcommand{\tab}[1]{\hyperref[tab:#1]{Table~\ref*{tab:#1}}}

\newcommand{\shrt}[1]{\hspace*{1pt}{#1}\hspace*{1pt}}
\newcommand{\?}{\hspace*{-0.5pt}}

\newcommand{\lt}[1]{_{\mathrm{#1}}}
\newcommand{\ltsm}[1]{_{\mathrm{\mathsmaller{#1}}}}

\newcommand{\ltn}[2]{_{\mathrm{#1},{#2}}}

\newcommand{\sast}{\hspace*{0.5pt}{\ast}\hspace*{0.5pt}}

\DeclareMathOperator{\E}{\mathbb{E}}

\newcommand{\ssumpt}[4]{%
	{\hspace*{-4pt}%
		\begin{array}{c}\scriptstyle{\mathclap{#2}}\\[#3]{\text{\larger[1]{$\sum$}}}\\[#4]\scriptstyle{\mathclap{#1}}%
		\end{array}\hspace*{-3pt}}}

\newcommand{\YY}{{Y\?Y}\!}

\copyrightnotice{978-1-6654-6867-1/22/\$31.00~\copyright2022 IEEE}
                   
\title{\vspace*{-1mm}
	Model-Based Estimation of In-Car-Communication Feedback\\
	applied to Speech Zone Detection%
\vspace*{-1.5mm}
}

\name{
	Kaspar M\"uller$^1$%
  	\thanks{This project has received funding from the SOUNDS European Training Network, an European Union’s Horizon 2020 research and innovation programme under the Marie Sk\l{}odowska-Curie grant agreement No.\ 956369.},
  	Simon Doclo$^2$,
  	Jan \O{}stergaard\,$^3$,
  	Tobias Wolff\,$^1$\vspace*{-1.5mm}%
  }
\address{
	$^1$Cerence GmbH, Acoustic Speech Enhancement, Ulm, Germany, {\normalsize\?\texttt{kaspar.mueller@cerence.com}}\\
	$^2$University of Oldenburg, Department of Medical Physics and Acoustics and\\ Cluster of Excellence Hearing4all, Oldenburg, Germany\\
	$^3$Aalborg University, Department of Electronic Systems, Aalborg, Denmark
	\vspace*{-3mm}%
}

\begin{document}

\maketitle
\ninept

\setlength{\abovedisplayskip}{6pt}
\setlength{\belowdisplayskip}{6pt}
\medmuskip=3mu

\begin{abstract}
Modern cars provide versatile tools to enhance speech communication.
While an in-car communication (ICC) system aims at enhancing communication between the passengers by playing back desired speech via loudspeakers in the car,
these loudspeaker signals may disturb a speech enhancement system required for hands-free telephony and automatic speech recognition.
In this paper, we focus on speech zone detection, i.e.\ detecting which passenger in the car is speaking, which is a crucial component of the speech enhancement system.
We propose a model-based feedback estimation method to improve robustness of speech zone detection against ICC feedback.
Specifically, since the zone detection system typically does not have access to the ICC loudspeaker signals, the proposed method estimates the feedback signal from the observed microphone signals based on a free-field propagation model between the loudspeakers and the microphones as well as the ICC gain.
We propose an efficient recursive implementation in the short-time Fourier transform domain using convolutive transfer functions.
A realistic simulation study indicates that the proposed method allows to increase the ICC gain by about $6$\,dB while still achieving robust speech zone detection results.
\end{abstract}
\vspace*{-1pt}

\section{Introduction}
\label{sec:intro}
\begin{figure}[tb]
	\centering
	\def\svgwidth{0.7\linewidth}
	{
		\sffamily\footnotesize
		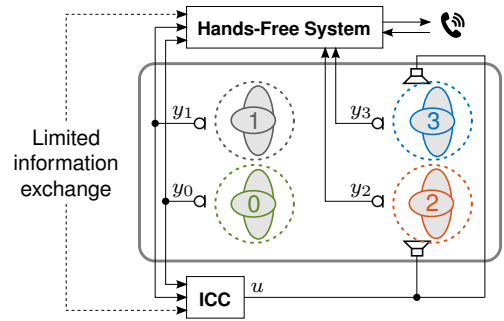\hspace*{5mm}
	}
	\vspace*{-6pt}%
	\caption{Exemplary car setup with independent hands-free and in-car communication\,(ICC) systems.\,Dashed circles symbolize speech zones.}
	\label{fig:car_setup}
	\vspace*{-3mm}
\end{figure}

\vspace*{-3pt}%
While multiple built-in loudspeakers for passenger entertainment in cars have been a standard for decades, modern car cabins are increasingly equipped with multiple distributed microphones for applications such as hands-free telephony, automatic speech recognition or in-car communication (ICC).
The former applications are designed for communication with external parties and require speech enhancement, e.g. beamforming or noise reduction \cite{Benesty2012,Buck2010,Matheja2013}.
On the other hand, ICC systems are designed to enhance speech intelligibility between passengers by reinforcing desired speech signals in the car cabin \cite{Ortega2002,Schmidt2006,Lueke2013}.
Often, this is achieved by recording the speech signal of a front passenger and reproducing it over loudspeakers at the rear cabin (see \fig{car_setup}) to prevent the driver from turning around or shouting in order to be understood.
The main challenge of ICC systems is to stabilize the closed electroacoustic loop resulting from the feedback of the loudspeaker signals to the microphones \cite{Ortega2002,Schmidt2006,Lueke2013,Waterschoot2011,Bulling2016,Bulling2017,Gimm2021}.
In practice, hands-free and ICC systems often run on different processors 
with highly restricted information exchange meaning that the loudspeaker signals used for the ICC system are generally not available for the hands-free system,
which rules out a joint processing of both systems as proposed in \cite{Gimm2020}.

\noindent
Depending on the ICC gain, the speech enhancement performance of the hands-free system may be substantially degraded by the ICC system.
In this paper, we specifically focus on its influence on speech zone detection:
When speech zones are defined in the car (one zone for each seat) to achieve speech enhancement for each zone individually, it is required to distinguish which zone is active, i.e.\ which passenger is speaking.
According to \cite{Matheja2017}, speech zone detection can be achieved by evaluating the maximum signal power ratios of passenger-dedicated microphones, as the speaker-dedicated microphone typically shows the highest signal power.
However, this assumption may be violated in combination with an ICC system since the signal power of microphones close to ICC loudspeakers may exceed that of the speaker-dedicated microphone.
One might consider classical feedback cancellation techniques \cite{Waterschoot2011} to remove the ICC feedback from the microphone signals.
This would however require a loudspeaker reference signal, which in practice is not available for speech zone \nolinebreak detection.

In this paper, a model-based feedback signal estimation method is proposed.
This method estimates the ICC feedback contribution in the observed microphone signals without requiring access to the clean loudspeaker signals and by only considering free-field propagation between the loudspeakers and microphones.
We propose an efficient recursive implementation in the short-time Fourier transform (STFT) domain using convolutive transfer functions (CTF) \cite{Talmon2009}.
Finally, we suppress the ICC feedback contribution from the power spectral densities (PSD) of the observed microphone signals, which helps to improve robustness of speech zone detection against ICC feedback.

\section{Signal Model}
\label{sec:signal_model}
\vspace*{-2pt}%

\begin{figure}[tp]
	\centering
	\def\svgwidth{0.95\linewidth}
	{
		\sffamily\footnotesize
		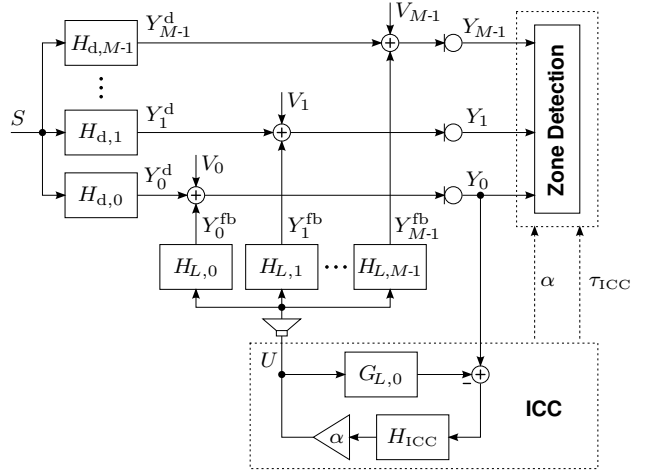
	}
	\vspace*{-5pt}%
	\caption{Signal model with basic in-car communication (ICC) system in the STFT domain (frequency and frame indices are omitted).}
	\label{fig:signal_model}
	\vspace*{-5pt}
\end{figure}

We consider a car environment with a single speaker, $M$ passenger-dedicated microphones and an ICC system with $L$ loudspeakers (see \fig{signal_model}).
The observed microphone signals $y_m(n)$, with the microphone index $m\shrt{=}\{0,...,M{-}1\}$ and the sample time index $n$, consist of three components:
The direct signals $y_m^\mathrm{d}(n)$, i.e.\ the microphone signals due to in-car speech (including reverberation), the ICC feedback signals $y_m^\mathrm{fb}(n)$ induced by the ICC loudspeakers as well as noise $v_m(n)$, i.e.\
\begin{align}
	y_m(n) &= y_m^\mathrm{d}(n) + y_m^\mathrm{fb}(n) + v_m(n)\,.
	\label{eq:noisy_y}
\end{align}
This paper will only consider the low-noise case, where $v_m(n)$ can be neglected.
The desired direct signal $y_m^\mathrm{d}(n)$ is the dry speech signal $s(n)$ filtered with the acoustic impulse response $h\ltn{d}{m}(n)$ from the speech source to the $m$-th microphone.
Likewise, the feedback signal~$y_m^\mathrm{fb}(n)$ can be expressed as
\begin{align}
	\!y_m^\mathrm{fb}(n) = u(n) \ast h_{\?L,m}(n) \;\text{with }  h_{\?L,m}(n) \shrt{=} 
	\textstyle \sum_{i=0}^{L-\?1} h_{i,m}(n),
	\label{eq:Lfilter}
\end{align}
where $u(n)$ denotes the ICC loudspeaker signal (same for all loudspeakers), $\{\ast\}$ denotes convolution and $h_{i,m}(n)$ is the acoustic impulse response from the $i$-th loudspeaker to the $m$-th microphone.
$m\shrt{=}0$ refers to the ICC reference microphone channel that is reinforced in the car cabin.

The ICC and speech enhancement system including the speech zone detection are implemented in the STFT domain. Accordingly, the signal model needs to be formulated in the same domain.
Since the impulse responses $h\ltn{d}{m}(n)$ and $h_{\?L,m}(n)$ are typically longer than the STFT frame size, and especially due to the recursive ICC structure (see \fig{signal_model}), filtering in the STFT domain requires a multi-frame approach such as crossband filtering \cite{Avargel2007,Avargel2008}.
To reduce computational complexity, we will consider the convolutive transfer function (CTF) approximation \cite{Talmon2009}, which only considers band-to-band filters, i.e.\
\vspace*{-5pt}
\begin{align}
	Y(k,l) &= S(k,l) \ast H(k,l) = 
	\ssumpt{l'\!=0}{N_{\?H}-1}{-0pt}{-0.5pt}
	S(k,l{-}l') \, H(k,l')
	\,,
	\label{eq:ctf_filtering}
\end{align}

\vspace*{-5pt}
\noindent
where $k$ denotes the frequency index and $l$ the frame index,
$S(k,l)$ and $Y(k,l)$ are the STFT representations of the input and output signal, respectively, and $H(k,l)$ is the CTF consisting of $N_{\?H}$ filter coefficients (we only consider causal coefficients).
Here, $\{\ast\}$ denotes the convolution over frames.


In this work, we model the ICC system (see \fig{signal_model}) using a linear model, where it should be realized that this is an approximation of a real ICC system, such as e.g.\ \cite{Ortega2002,Schmidt2006,Lueke2013}.
The forward path incorporates the static ICC CTF $H\ltsm{ICC}(k,l)$, which models the known ICC processing delay $\tau\ltsm{ICC}$, as well as a dynamic, real-valued and frequency-dependent ICC gain $\alpha(k,l)$.
Since $\tau\ltsm{ICC}$ is fixed, the ICC gain $\alpha(k,l)$ is the only variable parameter that needs to be transmitted from the ICC to the zone detection at runtime.
However, the ICC gain is usually slowly time-varying and thus requires no frame-wise transmission (the frame index will be omitted below).
$G_{\?L,0}(k,l)$ is a linear model of the ICC feedback cancellation filter (see \cite{Schmidt2006,Lueke2013,Bulling2016,Bulling2017,Gimm2021} for practical examples),
which aims at subtracting the feedback signal from the ICC input signal to stabilize the ICC system and only amplify the desired direct signal.

For better readability, the frequency and frame indices $k,l$ are omitted hereinafter in this section.
Using this signal model, it can be shown that the loudspeaker signal $U$ and the $m$-th feedback signal $Y_m^\mathrm{fb}$ are given by
\vspace*{-2pt}
\begin{gather}
	U = \alpha\,H\ltsm{ICC} \ast ( Y_0 - G_{\?L,0} \ast U )\,,
	\label{eq:U}
	\\
	Y_m^\mathrm{fb} = H_{\?L,m} \ast U\,.
	\label{eq:Yfb}
\end{gather}
By substituting \eqref{eq:U} into \eqref{eq:Yfb}, the feedback signal can be written as
\begin{align}
	\boxed{
	Y_{m}^{\mathrm{fb}} = 
	\alpha\,H\ltsm{ICC} \ast \big( H_{\?L,m} \ast Y_0 - G_{\?L,0} \ast Y_{m}^\mathrm{fb} \big).
	}
	\label{eq:Yfb_Y0}
\end{align}
We consider only strictly causal CTFs (exclusively depending on past frames) to ensure that the recursive filters in \eqref{eq:U}, \eqref{eq:Yfb_Y0} and in the following equations are realizable.

\noindent
To get a deeper understanding of the influence of the feedback cancellation filter $G_{L,0}$, one can define a mismatch
\begin{align}
	\Delta H_{\?L,0} &= H_{\?L,0} - G_{\?L,0}
	\label{eq:mismatch}
\end{align}
between the acoustic transfer function $H_{\?L,0}$ and $G_{L,0}$.
Resolving \eqref{eq:mismatch} for $G_{L,0}$ and inserting into \eqref{eq:U} yields
\begin{align}
	\begin{aligned}
		U &= 
	\alpha\,H\ltsm{ICC} \ast \big( Y_0 - H_{\?L,0} \ast U + \Delta H_{\?L,0} \ast U \big)
	\\
	&= 
	\alpha\,H\ltsm{ICC} \ast \big( Y_0 - Y_0^\mathrm{fb} + \Delta H_{\?L,0} \ast U \big).
	\end{aligned}
\label{eq:U2}
\end{align}
By using $Y_0 \,\shrt{=}\, Y_0^\mathrm{d} + Y_0^\mathrm{fb}$ and substituting \eqref{eq:U2} into \eqref{eq:Yfb}, the feedback signal can be written as
\begin{align}
	\boxed{
	Y_m^\mathrm{fb} = 
	\alpha\,H\ltsm{ICC} \ast \big( H_{\?L,m} \ast Y_0^\mathrm{d} + \Delta H_{\?L,0} \ast Y_m^\mathrm{fb} \big).
	}
	\label{eq:Yfb_Yd}
\end{align}
According to this, the mismatch $\Delta H_{L,0}$ can be interpreted as a measure of the amount of feedback reduction:
$\Delta H_{L,0} \,\shrt{=}\, 0$, i.e.\ $G_{\?L,0}\,\shrt{=}\,H_{\?L,0}$, corresponds to perfect feedback cancellation (only the desired direct signal $Y_0^\mathrm{d}$ would be amplified since there would be no recursion in \eqref{eq:Yfb_Yd});
$\Delta H_{\?L,0}\shrt{=}H_{\?L,0}$, i.e.\ $G_{\?L,0}\shrt{=}0$, represents no feedback cancellation.

\begin{figure*}[tp]
	\centering
	\def\svgwidth{0.73\linewidth}
	{
		\sffamily
		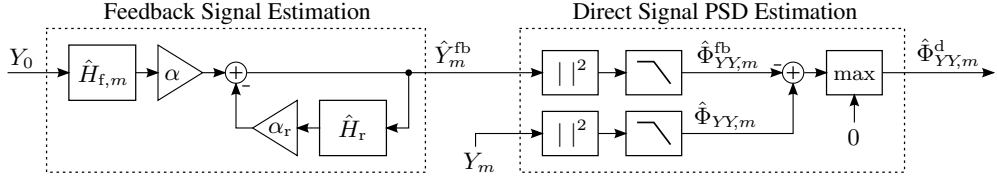
	}
	\vspace*{-5pt}%
	\caption{Signal flow graph of the direct signal PSD estimation in the STFT domain (frequency and frame indices are omitted).}
	\label{fig:DPE}
	\vspace*{-6pt}
\end{figure*}

\section{Feedback Signal Estimation}
\label{sec:feedback_signal_estimation}
\vspace*{-2pt}%

In practice, the feedback signals have to be estimated based on the observed microphone signals $Y_m(k,l)$ without knowledge of the loudspeaker signal $U(k,l)$.
For this purpose, a feedback signal estimator can be derived from \eqref{eq:Yfb_Y0} as
\begin{align}
	\begin{aligned}
		\hat Y_m^\mathrm{fb}(k,l) =\;
		&\,\alpha(k) \, \hat{H}\ltn{f}{m}\?(k,l) \ast {Y}_{0}(k,l) \;-
		\\ 
		&\alpha\lt{r}\?(k) \, \hat{H}\lt{r}\?(k,l) \ast \hat{Y}_m^\mathrm{fb}(k,l)\,,
	\end{aligned}
	\label{eq:Yfb_estimation}
\end{align}
where $\hat H\ltn{f}{m}\?(k,l) \,\shrt{=}\, {H}\ltsm{ICC}\?(k,l) \sast \hat H_{\?L,m}(k,l)$ denotes the feed-forward CTF and $\hat H\lt{r}\?(k,l) \,\shrt{=}\, {H}\ltsm{ICC}\?(k,l) \ast \hat G_{\?L,0}(k,l)$ the recursive CTF of the feedback signal estimator, respectively (see \fig{DPE}), and $\{\hat{\cdot}\}$ denotes estimated or modeled quantities.
We consider only strictly causal coefficients of the CTFs $\hat{H}\ltn{f}{m}\?(k,l)$ and $\hat{H}\lt{r}\?(k,l)$.
Furthermore, we introduced $\alpha\lt{r}(k)$ to ensure stability of the estimator, as~described~below.

According to \cite{Avargel2007,Talmon2009}, the CTFs $\hat H\ltn{f}{m}\?(k,l)$ and $\hat H\lt{r}\?(k,l)$ can be computed from the corresponding time-domain filters
\begin{align}
	\!\!\hat h\ltn{f}{m}(n) = h\ltsm{ICC}(n) \sast \hat h_{\?L,m}(n) ,\;
	\hat h\lt{r}(n) = h\ltsm{ICC}(n) \sast \hat g_{\?L,0}(n) \,.
	\label{eq:h_rf_hat}
\end{align}
The ICC transfer function $h\ltsm{ICC}(n)$ incorporates the known ICC processing delay $\tau\ltsm{ICC}$.
The acoustic impulse response $\hat h_{\?L,m}(n)$ can be either measured or modeled.
In this work, we consider a simple free-field propagation model.
Accordingly, the acoustic impulse response from the $i$-th loudspeaker to the $m$-th microphone are approximated\,by
\begin{align}
	\hat h_{i,m}(n) &= \frac{\beta_{i,m}}{D_{i,m}} \, \delta(n\shrt{-}\tau_{i,m}) \;
	\text{ with }
	\tau_{i,m} = \Big\lfloor \frac{\mathsmaller{f_s \,\cdot\, D_{i,m}}}{\mathsmaller{c}} \Big\rfloor,
	\label{eq:H_i_model}
\end{align}
where $D_{i,m}$ is the distance between the $i$-th loudspeaker and $m$-th microphone, $\beta_{i,m}$ is a gain factor that incorporates the individual sensitivity of the loudspeakers and microphones%
\footnote{$\beta_{i,m}$ reflects the level difference between the loudspeaker and microphone signal that would be observed in a free field at $D_{i,m}\shrt{=}1$m. It could also include the directivity of loudspeakers and microphones which is not considered here.}.
Relative to \eqref{eq:mismatch}, we propose to model the time-domain feedback cancellation filter as
\begin{align}
	\hat g_{\?L,0}(n) = (1-\lambda)\, \hat h_{\?L,0}(n)
	\label{eq:mismatch_model}
\end{align}
with $\hat h_{\?L,0}(n)$ as described above.
The mismatch factor $\lambda \shrt{\in} [0,1]$ introduced herein is used to model the amount of feedback reduction, where higher values of $\lambda$ correspond to less feedback cancellation when assuming that $\hat h_{\?L,0}(n) \approx h_{\?L,0}(n)$.

In practice, it is crucial to ensure stability of the recursive filter in \eqref{eq:Yfb_estimation} at runtime, where it should be noted that stability of the ICC system does not guarantee stability of the feedback signal estimator.
For this purpose, the gain $\alpha\lt{r}(k)$ of the recursive CTF is upper-limited
\begin{align}
	\alpha\lt{r}(k) = \min \big\{ \alpha(k), \,\alpha\lt{r,max}(k) - \delta \big\}\,,
\end{align}
where $\alpha\lt{r,max}(k)$ characterizes the stability border of the recursive filter in \eqref{eq:Yfb_estimation} and $\delta$ is a small positive value (we used $\delta \shrt{=} 0.2$ in this work).
$\alpha\lt{r,max}(k)$ can be determined as the maximum positive, real-valued gain for each subband~$k$, where all poles of the recursive filter
\begin{align}
	\hat H_k(z) &= 1\, \big/ \left(
		1 + \textstyle \sum_{l=1}^{L\lt{H}} \alpha\lt{r,max}(k) \, \hat H\lt{r}(k,l) \, z^{-l}
		\right)
\end{align}
are strictly within the unit circle \cite{Oppenheim1999}.

\section{\?Zone Detection Robust Against ICC Feedback}
\label{sec:zone_detection_and_PSD}
\vspace*{-2pt}

In this section, we describe an energy-based speech zone detection approach and moreover propose a method to increase its robustness against ICC feedback using the feedback signal estimates of \sect{feedback_signal_estimation}.

\vspace*{-3pt}
\subsection{Energy-Based Speech Zone Detection}
\label{sec:zone_detection}
In \cite{Matheja2017}, it was proposed to perform energy-based speech speech zone detection based on the signal power ratios (SPR)\footnote{Originally, the authors used the SPR for speaker activity detection.}
\begin{align}
	{\operatorname{SPR}}_m(k,l) &= 10\,\log_{10}
	\frac{\max \big\{ \hat\Phi_{\YY\?,m}(k,l),\epsilon \big\}
	}{\max \Bigg\{
		\displaystyle \max_{\stackrel{m' \in \mathcal{M}}{\scriptscriptstyle m' \neq m}}
		\big\{  \hat\Phi_{\YY,m'}(k,l) \big\}, \epsilon
		\Bigg\}}
	\label{eq:SPR}
\end{align}
of passenger-dedicated microphones, where $\hat \Phi_{\YY\?,m}(k,l)$ is the estimated PSD of the $m$-th microphone signal, $\mathcal{M}$ is the set of all microphones and $\epsilon$ is a small positive number.
A broadband SPR value can be determined by averaging over a set of frequencies $k \shrt{\in} \mathcal{K_\text{speech}}$ assumed to contain speech energy
(we used $100$\,Hz\,...\,$8$\,kHz), i.e.\
\vspace*{-6pt}%
\begin{align}
	\overline{\operatorname{SPR}}_m(l) &= 
	\frac{1}{|\mathcal{K}_\text{speech}|}\quad\,
	\ssumpt{k \in \mathcal{K}_\text{speech}}{}{-2pt}{-2pt} \operatorname{SPR}_m(k,l)\,.
	\label{eq:SPR_broadband}
\end{align}
The active zone is finally identified as the passenger-dedicated (= zone-dedicated) microphone with the largest broadband SPR, i.e.\
\begin{align}
	\zeta_\text{active}(l) &= \arg \max_{m \in \mathcal{M}} \, \overline{\operatorname{SPR}}_m(l)\,,
	\label{eq:zone_detection}
\end{align}
where $\zeta \shrt{\in} \mathcal{M}$ are the possible speech zones.

Robustness of this speech zone detection approach against ICC feedback could be increased by using direct signal PSD estimates $\hat\Phi_{\YY\?,m}^\mathrm{d}(k,l)$ (no contribution of ICC feedback) instead of the microphone signal PSDs $\hat\Phi_{\YY\?,m}(k,l)$ to compute the SPR in \eqref{eq:SPR}.
Hereinafter, we propose an estimation of the direct signal PSDs using the feedback signal estimates from \sect{feedback_signal_estimation}.

\subsection{Estimation of the Direct Signal PSD}
\label{sec:direct_signal_PSD}

One could consider directly subtracting the estimated feedback signal $\hat Y_m^\mathrm{fb}(k,l)$ from the observed microphone signal $Y_m(k,l)$ to obtain a direct signal estimate $\hat Y_m^\mathrm{d}(k,l)$ in a first step.
However, this approach is highly sensitive to estimation errors such as phase errors.
Therefore we propose to estimate the direct signal PSD in the power domain, which enables a better control of estimation errors.
The microphone signal PSD $\Phi_{\YY\?,m}(k,l)$ is defined as
\begin{align}
	\begin{aligned}
		\!\E \?\big\{ |Y_m(k,l) |^2 \big\}
	=\,&\E \?\big\{ | Y_m^\mathrm{d}(k,l) |^2 \big\} +
	\E \big\{ | Y_m^\mathrm{fb}(k,l) |^2 \big\} \,+
	\\
	& \,2 \, \Re \big\{ \?\E \?\big\{ Y_m^\mathrm{d}(k,l) \, Y_m^{\mathrm{fb}*}(k,l) \big\} \!\big\}\,,
	\end{aligned}
	\label{eq:PSD_Y}
\end{align}
where $\E\{ \cdot \}$ denotes the expectation operator and $\Re\{ \cdot \}$ the real value operator.
It can be shown that the cross term 
$\E \big\{ Y_m^\mathrm{d}(k,l) \, Y_m^{\mathrm{fb}*}(k,l) \big\}$ is negligible
if the STFT frame size $N \shrt{<} \tau\ltsm{ICC} \shrt{+} \min_i \{\tau_{i,m}\}$ (see~\eqref{eq:H_i_model})
and the source signal $S(k,l)$ is a zero-mean random signal.
Moreover, this condition implicitly entails that the CTFs $\hat H\lt{r}\?(k,l)$ and $\hat H\ltn{f\?}{m}\?(k,l)$ in \eqref{eq:Yfb_estimation} are strictly causal, which ensures the realizability of the feedback signal estimator.

\!\!\!Real-life scenarios in the car do not fully comply with the conditions stated above as speech signals are short-time stationary and reverberation is ignored.
However, assuming low reverberation times in a car and ignoring short-time stationary of speech, we propose to estimate the direct signal PSD (complying with the upper bound of $N$) as
\begin{align}
	\hat\Phi_{\YY\?,m}^\mathrm{d}(\?k,\?l) &= 
	\max \big\{
	\hat\Phi_{\YY\?,m}(\?k,\?l) - \hat\Phi_{\YY\?,m}^\mathrm{fb}(\?k,\?l), 0
	\big\}.
	\label{eq:dspsd_model}
\end{align}
Here, the direct signal PSD estimate $\hat\Phi_{\YY\?,m}^\mathrm{d}(\?k,\?l)$ is lower-limited to avoid negative results due to estimation errors.
At runtime, the right-hand side PSD estimates in \eqref{eq:dspsd_model} are computed by exponential smoothing of the squared magnitudes of the according spectra ${Y}_m(k,l)$ and $\hat{Y}_m^\mathrm{fb}(k,l)$, respectively (see \fig{DPE}).

\section{Simulation Results}
\label{sec:results}
\vspace*{-2pt}%
The influence of the proposed direct signal PSD estimation on the described speech zone detection approach was evaluated in simulations.

\vspace*{3pt}
\noindent
\textit{Microphone signal simulation:}\quad
The time-domain microphone signals were simulated using the signal model described in \sect{signal_model} at a sampling frequency $f_s\shrt{=}24$\,kHz with a clean female speech source sample.
The acoustic impulse responses $h\ltn{d}{m}(n)$ and $h_{\?L,m}(n)$ were measured in a car (Audi A6) with reverberation time $T_{60} \shrt{\approx} 80$\,ms using $M\shrt{=}4$ passenger-dedicated microphones as illustrated in \fig{car_setup}.
The direct impulse responses $h\ltn{d}{m}(n)$ were measured with a GRAS 44AA mouth simulator at zone 0 (driver seat), while $h_{\?L,m}(n)$ were measured with $L\shrt{=}2$ built-in rear door loudspeakers.
Uncorrelated white noise was added at each microphone ($\operatorname{SNR}\shrt{=} 20$\,dB at the reference microphone).
To simulate the ICC system, we only considered a stationary broadband ICC gain $\alpha$ and
$h\ltsm{ICC}(n)$ consisted of a processing delay $\tau\ltsm{ICC} \shrt{=} 15$\,ms and a fixed gain.
The latter gain was adjusted so that the ICC system at $\alpha \shrt{=} 0$\,dB (typical ICC operation gain) induces the same RMS level at the rear microphones ($m \shrt{=} 2,3$) as at the co-driver microphone ($m\shrt{=}1$) due to a signal from the driver seat.
For the feedback cancellation filter $g_{\?L,0}(n)$ in the ICC system, the measured impulse response $h_{\?L,0}(n)$ was superimposed by a random, white noise mismatch $\Delta h_{L,0}(n)$ (cf.~\eqref{eq:mismatch}).
It was adjusted so that the simulated ICC system became unstable for $\alpha\shrt{>} 4$\,dB, which matches a realistic system.

\vspace*{3pt}
\noindent
\textit{Direct signal PSD estimation:}\quad
The modeled impulse responses $\hat h_{\?L,m}(n)$ and $\hat g_{\?L,0}(n)$ which are required for the feedback signal estimation (\sect{feedback_signal_estimation}) were defined as follows:
$\hat h_{\?L,m}(n)$ was modeled according to
\eqref{eq:H_i_model}
(free-field propagation), where the distances $D_{i,m}$ between the ICC loudspeakers and the passenger-dedicated microphones were measured in the car.
Furthermore, the same broadband gain $\beta_{i,m}$ was assumed for all impulse responses (same sensitivity for all loudspeakers and microphones is assumed).
The modeled feedback cancellation filter in \eqref{eq:mismatch_model} was set to $\hat g_{\?L,0}(n) \shrt{=} \hat h_{\?L,0}(n)$.
We intentionally chose this deviation from the simulated feedback cancellation filter $g_{\?L,0}(n)$ to investigate the robustness against modeling inaccuracies.
The feedback signal estimation and direct signal PSD estimation was implemented in the STFT domain with Hann-windowed frames of $N\shrt{=}256$ samples length (corresponding to $10.7$\,ms) and $50\%$ overlap.
\\
\vspace*{1pt}

\begin{figure}[tb]
	\centering
	\centerline{\includegraphics[width=0.97\linewidth]{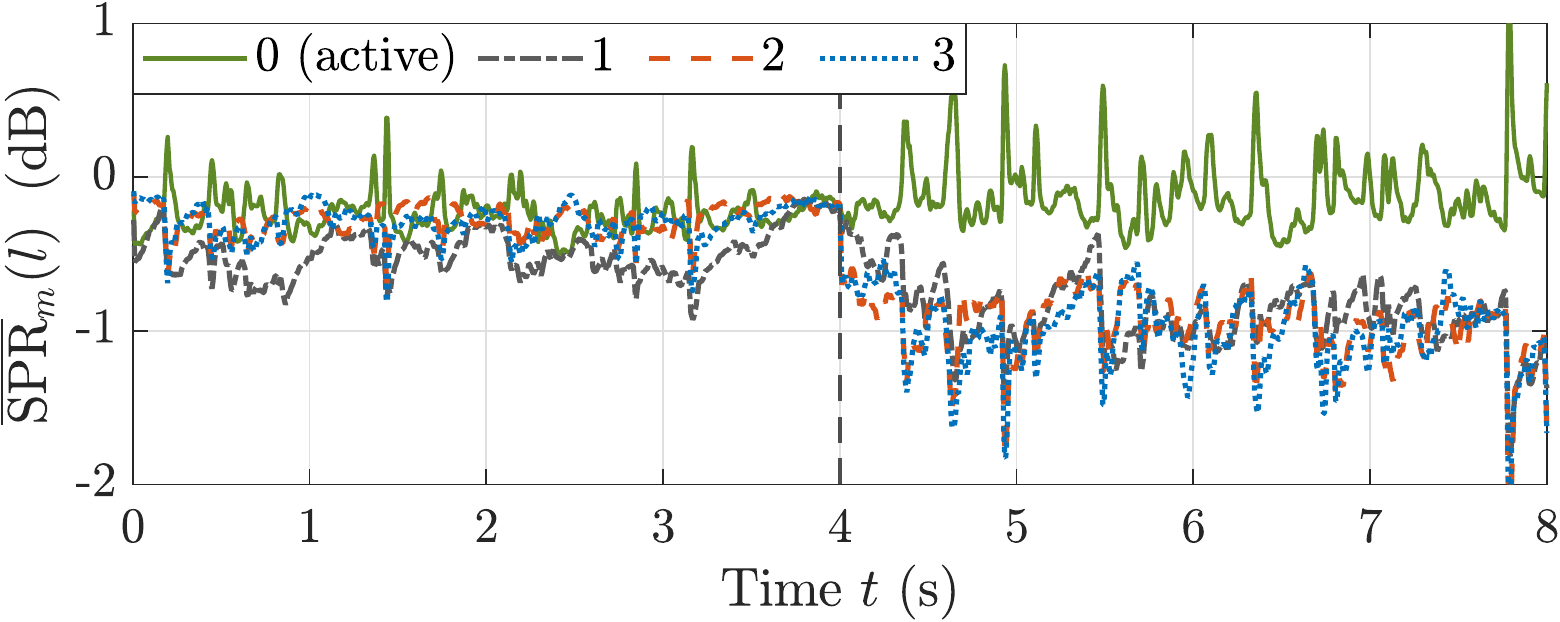}}
	\vspace*{-10pt}%
	\caption{Broadband SPR levels for an 8s speech signal at ICC gain $\alpha \shrt{=} 0$\,dB based on the unprocessed microphone signal PSDs (0s-4s) and on the proposed direct-sound PSD estimates (4s-8s).}
	\label{fig:results_SPR0}
	\vspace*{10pt}
	\begin{minipage}[b]{1.0\linewidth}
		\centering
		\centerline{\includegraphics[width=0.97\linewidth]{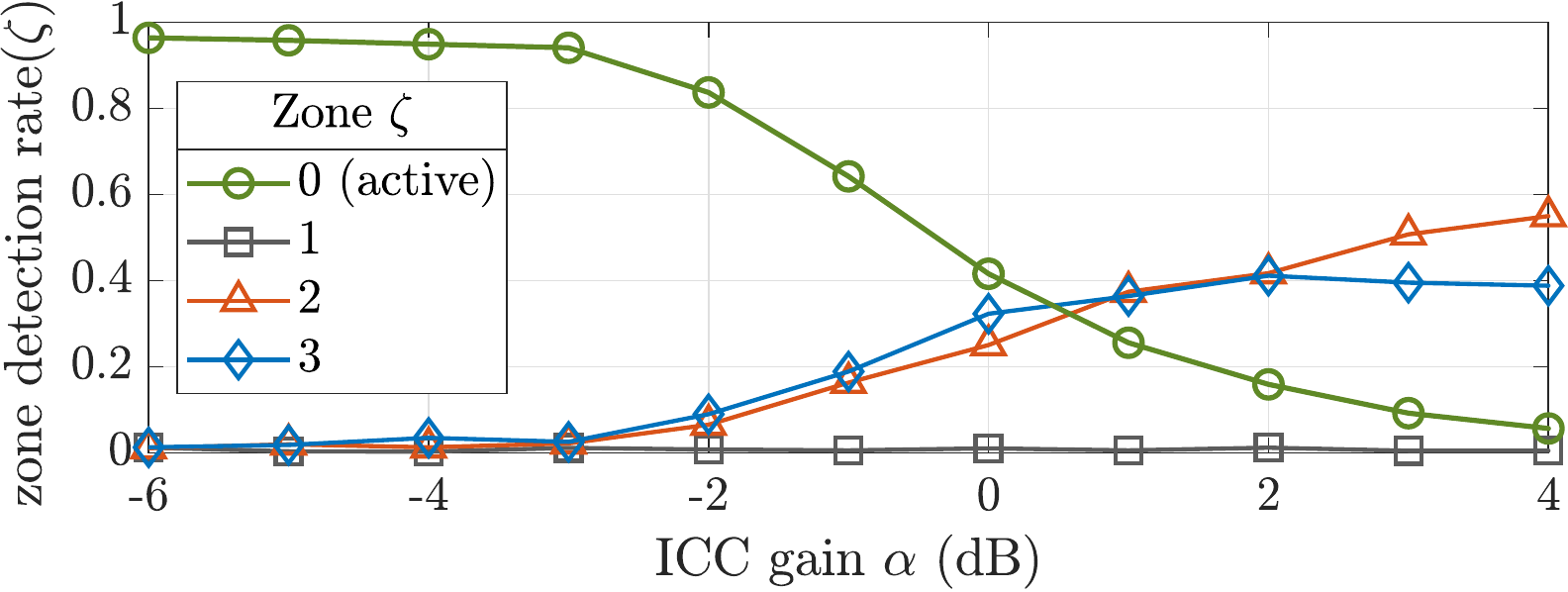}}
		{(a) Zone detection rates using unprocessed microphone signal PSDs.}\medskip
	\end{minipage}
	
	\vspace*{1pt}
	\begin{minipage}[b]{1.0\linewidth}
		\centering
		\centerline{\includegraphics[width=0.97\linewidth]{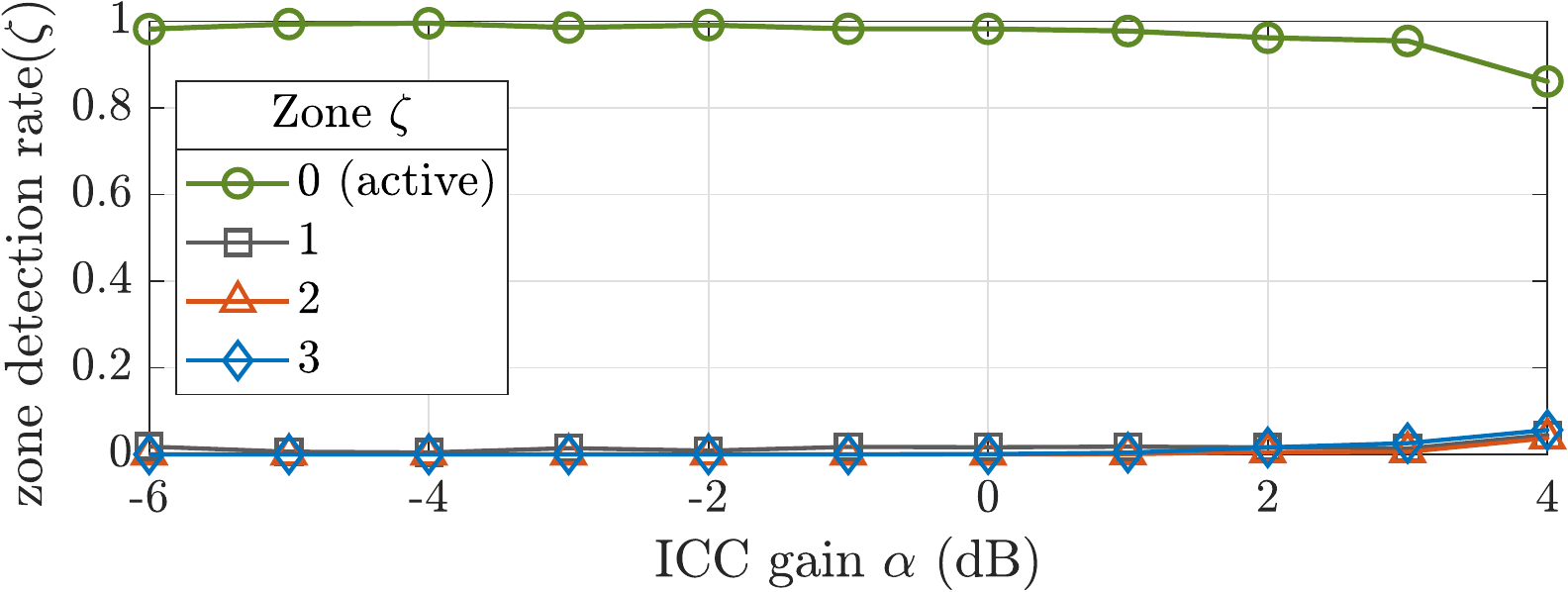}}
		{(b) Zone detection rates using the proposed direct signal PSDs.
		}
	\end{minipage}
	\vspace*{-14pt}
	\caption{Zone detection rates of an 8s speech signal in zone~$\zeta \shrt{=}0$ (driver seat) for different ICC gains.
	$\operatorname{zone\;detection\;rate}(0)$ (circle markers) corresponds to the rate of correct detection.}
	\label{fig:results_ZDR}
	\vspace*{-2.5mm}
\end{figure}

\noindent
\textbf{Results and Discussion}
\vspace*{6pt}

\noindent
\fig{results_SPR0} shows the broadband SPR in \eqref{eq:SPR_broadband} of the four passenger-dedicated (= zone-dedicated) microphones for an 8s speech signal from the driver seat (zone 0) at a typical ICC gain $\alpha \shrt{=} 0$\,dB.
The proposed direct signal PSD estimation was activated after 4s.
As can be clearly observed, the SPR levels without processing (0s-4s) are largely overlapping whereas a complete separation between the SPR levels of the active zone 0 and the remaining zones is achieved by the proposed method (4s-8s), even though the mean SPR levels are shifted by less than $1$\,dB.

\fig{results_ZDR} shows the zone detection rate of the four speech zones
\begin{align}
	\operatorname{zone\;detection\;rate}(\zeta) &= \frac{\# \operatorname{frames\, with } \zeta_\text{active}(l)\shrt{=}\zeta}{\# \operatorname{total\;frames}}
\end{align}
for different ICC gains $\alpha$, where the same 8s speech signal as before was used.
The $\operatorname{zone\;detection\;rate}(0)$ (circle markers) thus reveals the rate of correct zone detection.
The speech zone detection either directly used the unprocessed microphone signal PSDs (a) or the estimated direct signal PSDs (b).
\fig{results_ZDR}a shows that the correct zone detection rate based on the microphone signal PSDs \cite{Matheja2017} is degraded to less than 50\% due to ICC feedback at a typical ICC gain $\alpha\shrt{=}0$\,dB.
In contrast, the results in \fig{results_ZDR}b indicate a substantial improvement due to the proposed direct signal PSD estimation, where the ICC-gain can be increased by about $+6$\,dB to obtain similar detection rates.
Moreover, good speech zone detection results are obtained until the stability limit of the simulated ICC system at $\alpha \shrt{=} 4$\,dB.

\section{Conclusions}
\label{sec:conclusions}
\vspace*{-3pt}%
This work described an approach to enhance the robustness of energy-based speech zone detection against an independently operating, interfering in-car communication system.
The proposed method was designed in particular to cope with very limited information exchange between the speech zone detection and the ICC system.
Specifically, we introduced a model-based ICC feedback signal estimation based on a free-field propagation model between loudspeakers and microphones, which requires no clean loudspeaker reference signal but only the slowly time-varying ICC gain.
A computational efficient implementation in the STFT domain was derived consisting of one feed-forward and one recursive CTF.
The resulting feedback signal estimates were used to estimate the microphone signal PSDs without ICC feedback.
Simulations with measured impulse responses in a car indicated a robustness gain of about $6$\,dB against ICC feedback.

While this work considered low-noise scenarios with a simulated ICC system and speech with frontal head orientation, future work should also focus noisy environments with a real, more complex ICC system and different head orientations of a speaker.

\pagebreak

\balance
\bibliographystyle{dep/IEEEbib}


\end{document}